# A multi-agent AI classroom based on dual-process reasoning hazards: a pilot with prospective physics teachers

Eugenio TUFINO

*Department of Physics, Informatics and Mathematics, University of Modena and Reggio Emilia, Via G. Campi, 213/A, Modena, Italy*

*E-mail: eugenio.tufino@unimore.it

**Abstract**

Responding productively, in real time, to the reasoning students actually produce is among the most difficult and slowest to acquire of the skills physics teachers develop, and prospective teachers get few opportunities to practise it before entering a classroom. In this pilot study, we describe a simulated class of five AI students, each consistently enacting a distinct reasoning pattern drawn from the dual-process theory (DPT) "hazards" framework of physics education research, and report on its first use in a university course for prospective physics teachers. Fifteen graduate students worked through an instructional sequence on DPT and questioning strategies. They then diagnosed two parallel sets of written vignettes, in a crossover arrangement, before and after interacting in pairs with the simulated class on a static-friction problem. The data set is completed by individual written hypotheses and reflections, the session logs and a short feedback questionnaire. Diagnostic scores improved significantly from PRE to POST ($n = 11$ paired, Wilcoxon $p = 0.014$, $r = 0.79$; on a 0–18 scale). During the simulation itself, however, participants asked predominantly uniform guiding questions, and the DPT vocabulary they had been taught appeared in only 2 of 71 substantive teacher turns — while seven of thirteen POST sheets used it readily. We read this knowing–doing gap not as a failure but as a snapshot of where each participant stands on the developmental trajectory of responsiveness to student ideas, a trajectory the simulation makes visible with transcript-level granularity. We discuss design, findings and implications for physics teacher preparation.

# 1. Introduction

.

Generative artificial intelligence (AI) has entered physics education mainly as a support for the learner. Large language models (LLMs) have been used as tutors and virtual teaching assistants [1,2], as guides during inquiry-based activities [3], as customisable Socratic dialogue partners [4], and as engines with which students and teachers build and refine physics simulations [5,6]; for an overview of this rapidly growing area, see the recent rapid review [7] and the focused collection introduced in [8]. A smaller line of work reverses the roles: the AI plays the student, and the human practises teaching. Gregorcic and colleagues prompted ChatGPT to act as a physics student so that teachers could hone their Socratic questioning [9]; outside physics, GPTeach let novice teaching assistants run office hours with GPT-simulated students [10].

In this paper we take the second approach a step further. Rather than a single simulated student, we built a small simulated class: five AI agents, each consistently enacting a distinct pattern of reasoning drawn from dual-process theories of reasoning (DPT) as developed in physics education research [11,12]. To our knowledge, simulated students whose behaviour is grounded in a discipline-specific cognitive framework of this kind have not previously been reported in physics teacher preparation.

The development of the competence required in such an environment is known to be difficult. Dodlek, Planinsic and Etkina found that pre- and in-service physics teachers who successfully identified productive and problematic elements in students' written explanations nevertheless rarely built on that reasoning when responding to the students [13]; responding in real time is harder still and requires specific habits. Yet prospective teachers typically meet this demand for the first time in the practicum, with real students and real stakes. A simulated class offers a low-stakes setting for deliberate practice [14] much earlier in a teacher's trajectory — and, unlike a classroom, it leaves a complete transcript of every exchange available for analysis and feedback.

In this pilot study, we describe the simulation and report on its first use in a university course for prospective physics teachers enrolled in MSc programs in Physics and Mathematics at the University of Modena and Reggio Emilia. The simulation did not stand alone: it completed an instructional sequence that moved from an introduction to DPT and to questioning strategies, through practice exercises and written diagnostic vignettes, to the simulated class and a collective debrief. We examine what participants did while interacting with the simulated class, and how their written diagnoses of student reasoning changed between a pre and a post diagnostic task administered one week apart. In section 2 we outline the theoretical background; section 3 presents the simulation; sections 4 and 5 describe methods and results; in sections 6 and 7 we discuss the findings and conclude.

# 2. Background

## 2.1 Dual-process theories and reasoning hazards

Dual-process theories (DPT) describe human reasoning as the interplay of two processes [15]. Process 1 is fast, automatic and intuitive: it produces a provisional mental model based on prior experience and contextual cues, and it cannot be switched off. Process 2 is slow, effortful and analytical: it evaluates the provisional model — if the model is judged satisfactory, it becomes the answer. Because engaging Process 2 costs effort, people tend to avoid it when they feel confident. Over the past decade, physics education research has drawn on these theories to explain a familiar puzzle: students who demonstrably possess the relevant knowledge still give incorrect answers to conceptual questions [16,17]. Kryjevskaia, Heron and Heckler mapped the points along the reasoning path at which such failures can occur, labelling them hazards A–D [12]; the framework has since guided instructional interventions and further empirical work [11,18].

Table 1 summarises the four hazards as they were presented in our course. In brief: intuition may supply an incorrect first model (A); the reasoner, feeling confident, may accept it without scrutiny — a manifestation of cognitive frugality (B); analysis may be engaged but biased towards rationalising the initial answer (C); or analysis may be engaged in good faith while the required knowledge, the mindware, is insufficient to detect the error (D).

**Table 1.** The four reasoning hazards [12], with the questioning strategy matched to each hazard in the course (see section 2.2). The same mapping underlies the design of the simulated students (section 3).

| Hazard | Where reasoning fails | Matched strategy |
|---|---|---|
| A | Process 1 produces an incorrect provisional model | Guiding question, to trigger Process 2 |

| B | Confident reasoner does not engage Process 2 (cognitive frugality) | (with A) Guiding question |
|---|---|---|
| C | Process 2 is engaged but biased: rationalisation confirms the initial model | Create a conflict between formula and known result (“What does your formula predict for this simpler case?”) |
| D | Process 2 is engaged in good faith, but mindware is insufficient | Falling-back question, to rebuild from familiar ground |

## 2.2 Questioning strategies

In the Investigative Science Learning Environment (ISLE) approach, students develop their knowledge of physics through processes that mirror those of practising physicists. Teachers use questions rather than explanations to guide this work, and are explicitly prepared to elicit students' ideas, listen to them, and choose questions that will move their reasoning forward. The ISLE approach distinguishes two main types of teacher questions [19]: guiding questions push the student forward to the next reasoning step and work when the student has the mindware but is not activating it; falling-back questions take the student back to a simpler, familiar situation from which understanding can be rebuilt. In the course, a third move complemented these for Hazard C, where an engaged but biased Process 2 rationalises the initial answer: questions that create a conflict between the student's formula and a known result (e.g. "What does your formula predict for this simpler case?"). Each hazard was explicitly matched to a strategy (table 1) in the lesson preceding the simulation.

## 2.3 Responding to student reasoning as a developing skill

Knowing a framework, however, is not the same as teaching with it. Dodlek, Planinsic and Etkina, investigating how pre- and in-service physics teachers respond to students' written explanations, found that participants who successfully identified productive and problematic elements of student reasoning "rarely built on student reasoning when responding to the students, mostly focusing on addressing problematic aspects" [13].

They note that responding in real time is harder still and requires specific habits. Encouragingly, the same study also shows that this competence responds to preparation: pre-service teachers trained in an ISLE-based programme were often comparable with, and at times better than, in-service teachers, a result the authors attribute in part to that preparation. This literature establishes an expectation for our study: prospective teachers should not be expected to demonstrate spontaneous, differentiated questioning during their first structured teaching experience. What matters is their position and what an instrument can reveal about it.

# 3. The instructional sequence and the simulation

## 3.1 The instructional sequence

The simulation was embedded in a four-lesson sequence (two academic hours each) within the course (Figure 1 shows the timeline and the activities).

**Lesson 1** introduced dual-process theories through first-hand experience. Participants answered individually, via an online polling tool, the two questions of the block-and-magnet sequence of Kryjevskaia et al [11]: a screening question (a block held by a rod on a table) and the target question (a magnet on a refrigerator, pushed upward by a hand). The class reproduced the typical pattern: 93% answered the screening question correctly, while the class split evenly (50%–50%) on the magnet question. Only then were the two reasoning processes, the four hazards and the terrain analogy introduced [12], so that participants encountered the framework as an explanation of their own reasoning, not as an abstract taxonomy.

**Lesson 2** completed the theoretical toolkit. After a brief treatment of the resources view of student knowledge [20], the four hazards were reviewed and the questioning repertoire of the ISLE framework was analysed from Etkina and Planinsic's guide [19]: open versus closed questions, wait time, and the distinction between guiding questions (pushing the student to the next reasoning step) and falling-back questions (returning to a simpler, familiar situation); a third move — creating a conflict between the student's formula and a known result — was matched to Hazard C, building on the block-and-magnet experience of Lesson 1. Participants then practiced in groups of three at whiteboards on written student answers taken from Etkina's teacher-preparation workshop materials and from the Active Learning Guide accompanying *College Physics: Explore and Apply* [21] (a bulb-ranking task and a circuit task; an elevator task was assigned as homework), identifying productive and problematic ideas and formulating one guiding and one falling-back question for each fictional student. The lesson closed with the PRE diagnostic vignettes (section 4). The instructional material is provided as supplementary material.

**Lesson 3** was devoted to the simulation session, described below. After a debrief of the elevator homework, participants completed the individual hypotheses sheet, interacted with the simulated classroom in pairs for approximately 30–40 minutes, exported the conversation log, and wrote the individual reflection.

**The first part of Lesson 4** closed the sequence: the instructor commented on excerpts from the simulation logs, revisited the agent–hazard mapping, and administered the POST vignettes and a short feedback questionnaire.

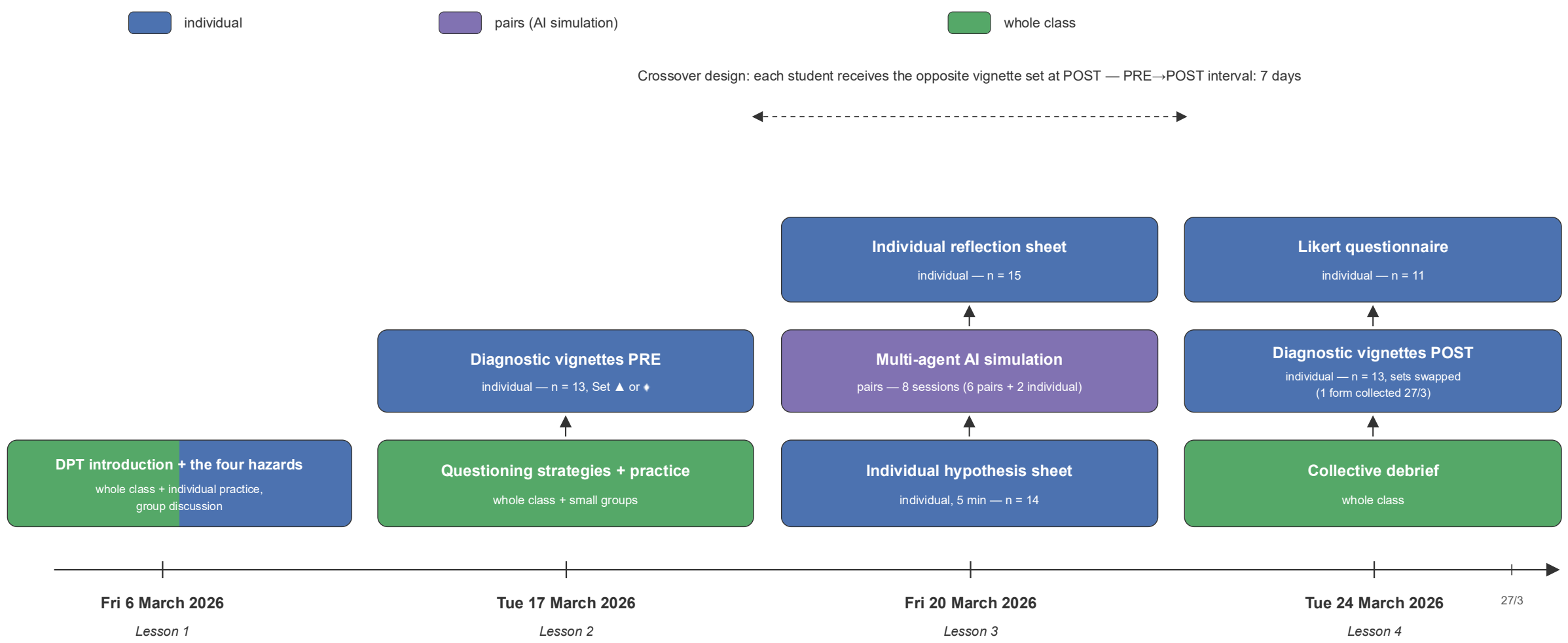

Figure 1 Timeline of the instructional sequence in the course.

## 3.2 The simulated classroom

The simulation presents a class of five AI students, each embodying a distinct point of failure in the reasoning process: Davide (Hazards A+B: an appealing wrong intuition, accepted without scrutiny), Elena (Hazard C: analytical reasoning recruited to defend the intuitive answer), Tommaso (Hazard D: insufficient knowledge), Chiara (a student in transition from intuitive to analytical reasoning) and Andrea (a near-expert peer). We stress that the hazards are failure points of a process, not student types; reifying them as stable characters is a deliberate pedagogical simplification, to which we return in the Discussion.

The class discusses a static-friction problem adapted from the opening vignette of Kryjevskaia, Heron and Heckler [12]: two identical boxes, B and C, rest on different surfaces ($\mu_s$ = 0.4 and 0.6); both are pulled with a horizontal force of 30 N and both remain at rest. Is the friction force on C greater than, less than, or equal to that on B? The correct answer — equal, 30 N on each, since $\mu_s N$ gives the maximum static friction, not the actual one — is designed so that Davide's intuition ("rougher surface, more friction") and Elena's formula-based rationalization converge on the same wrong answer, while the near-expert Andrea states the correct reasoning from the outset.

The participant, in the teacher's role, interacts with the class in written chat. An agent that is named responds; with roughly 30% probability a second agent intervenes; an open question draws two to four responses. The environment runs on a Large Language Model (LLM)[1] via a web interface, with session reset and one-click export of the anonymised conversation log (see Figure 2). The simulation is a Python application built with the Gradio library and hosted on Hugging Face Spaces platform. The five student personas are defined entirely in the system prompt, which specifies each agent's reasoning profile, typical language and degree of confidence.

The prompt was developed iteratively: we tested the agents' responses against the intended hazard profiles and refined the persona descriptions until the responses were pedagogically satisfactory. No specialised infrastructure is required, and the same application can be run with other LLM models supported by the Hugging Face platform. The application requires a valid API key from a supported LLM provider (to be supplied by the user) for operation. The simulation will be made openly available upon publication, so that teacher educators can try it or adapt it to their own scenarios by inserting their own credentials.

[1] Claude Sonnet 4.6 (Anthropic), through a Gradio application hosted on Hugging Face Spaces.

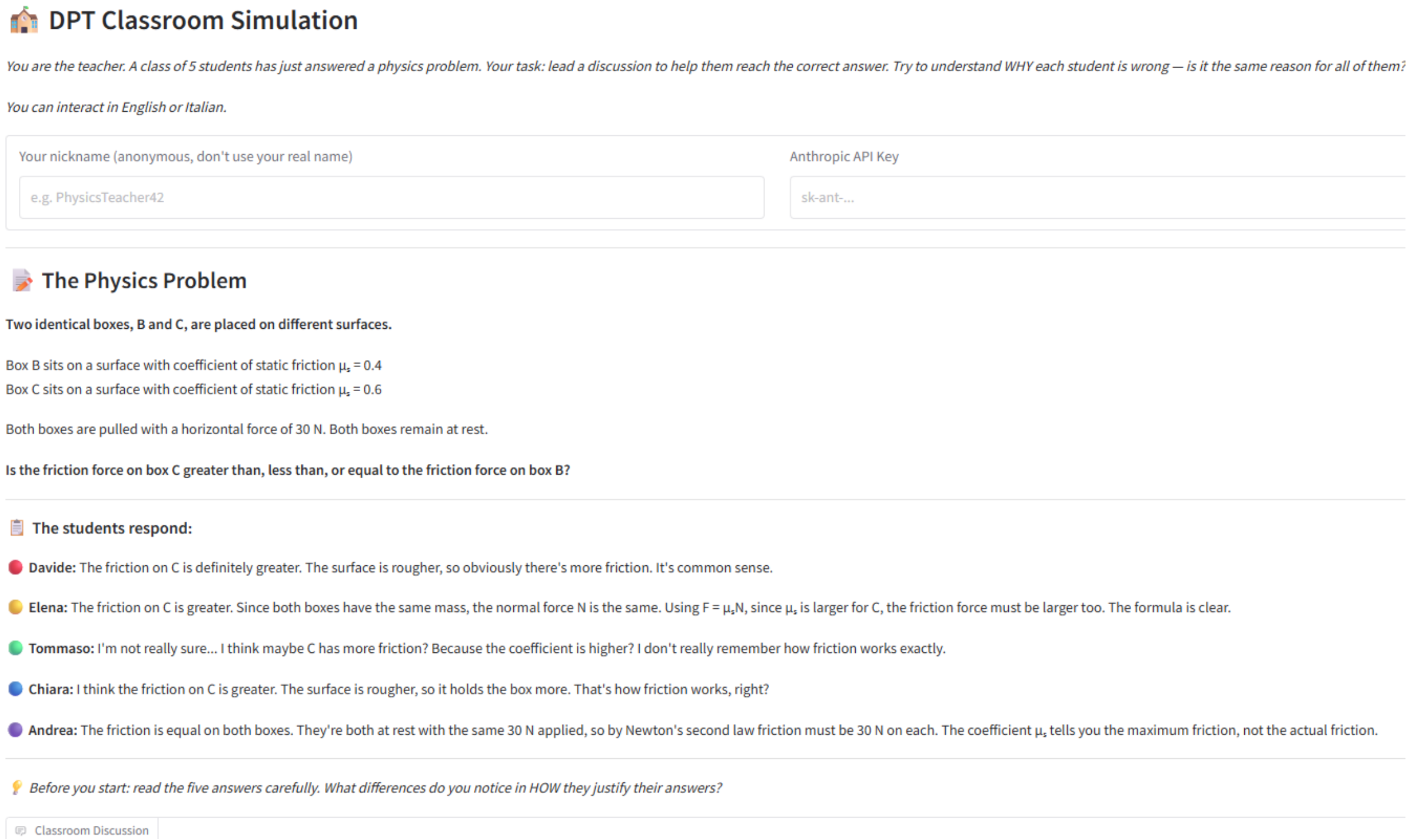

Figure 2. Screenshot of the DPT classroom simulation on Hugging Face platform. The problem and the initial responses from the five AI students are presented.

The following exchange, from the opening of one session, illustrates the interaction (all excerpts in this paper are verbatim from anonymized session logs, translated from Italian in some cases):

> ***Participant teacher:*** *Very well, let's begin: I saw your answers to the question and wanted to ask you a few things: for instance, Davide, do you think what you said holds both for objects at rest and for objects in motion?*
>
> ***Davide:*** *Yes, it always holds. Rougher surface, more friction. Whether it's at rest or moving doesn't change the basic idea.*
>
> ***Participant teacher:*** *Then why, in your view, are both boxes at rest if the net forces are different?*

Within three turns, the participant has elicited Davide's intuitive rule and set up a conflict with the observed situation — the kind of move whose distribution across sessions we examine in section 5.

# 4. Methods

## 4.1 Participants and context

Fifteen graduate students took part in the study as part of a Physics Education Research course at the University of Modena and Reggio Emilia (academic year 2025–26). The group was heterogeneous in background: some participants came from a mathematics education curriculum and had prior coursework in education, while for the others, enrolled in other physics curricula, this was their first education course; not all participants intend to pursue a teaching career. All participants gave written informed consent to participate and to the publication of anonymised data.

## 4.2 Design and data collection

The study followed a pre/post design built around a single simulation session (Figure 1). After instruction on the dual-process hazards and on ISLE questioning strategies, participants completed the PRE diagnostic vignettes. In the following lesson, each participant first completed an individual hypotheses sheet, predicting how the simulated students would reason; participants then interacted with the simulation in pairs (eight session logs: six pairs and two individual sessions), and immediately afterwards wrote an individual written. Working in pairs was both a practical and a pedagogical choice: it kept the number of simultaneous sessions low, and, in a classroom setting, it let each pair discuss and negotiate the next question before typing it.

In the following lesson, a whole-class debrief was followed by the POST vignettes and a five-item Likert questionnaire. The POST was therefore administered seven days after the PRE and four days after the simulation. Since the PRE itself followed the theoretical instruction, the measured gains capture what the simulation, the written reflections and the debrief added to theory alone, a combined effect that the design cannot separate. Of the fifteen participants, thirteen completed the PRE vignettes and thirteen the POST; twelve completed both (one took part in the simulation but in neither written task, and two others completed only one of the two). One of the twelve completed the PRE under uncontrolled conditions and was excluded from the paired analysis, which therefore comprises n = 11 participants; this participant's simulation log was retained in the qualitative analysis. One POST sheet, submitted three days late, was retained.

## 4.3 Instruments and scoring

Each vignette set (see figure 3) presents three fictional secondary-school students whose reasoning about a mechanics problem instantiates, respectively, Hazards A+B, C and D. Two parallel sets were used in a crossover arrangement — coins falling from a table (set ▲) and a mass–spring system taken to the Moon (set ♦) — with each participant receiving at POST the set not seen at PRE.

**Diagnostic Vignette — Set ▲**

Physics Education Course — University of Modena and Reggio Emilia

**Nickname:** ____________ **Date:** ____________

**Scenario.** Two identical coins are released simultaneously from the edge of a table. One coin is simply dropped; the other is launched horizontally with a flick. Air resistance is negligible. A student is asked: *"Which coin hits the ground first, or do they hit at the same time?"*

*For each of the following student responses, answer the two questions below.*

**Student 1 — Marco.** "The dropped coin hits the ground first — it goes straight down, while the other one has to travel a longer path through the air." When the teacher asks him to think about it more carefully, he says: "It makes sense — a longer path takes more time."

a) Why do you think Marco gave this answer?

b) What question would you ask Marco to help him?

**Student 2 — Sara.** "The dropped coin hits first. The launched coin has a total velocity $v = \sqrt{v_x^2 + v_y^2}$ which is larger than the dropped coin's velocity $v_y$. But it also covers a longer distance — the diagonal path $d = \sqrt{x^2 + h^2}$ is greater than the height $h$. Since it travels a longer distance, it takes more time even though it moves faster."

a) Why do you think Sara gave this answer?

b) What question would you ask Sara to help her?

**Student 3 — Luca.** "I'm not sure... maybe the launched coin takes longer? Or the same time? I think there's something about horizontal and vertical motions being separate, but I can't really explain how that works here."

a) Why do you think Luca gave this answer?

b) What question would you ask Luca to help him?

**Diagnostic Vignette — Set ♦**

Physics Education Course — University of Modena and Reggio Emilia

**Nickname:** ____________ **Date:** ____________

**Scenario.** A block attached to a spring oscillates vertically with a period $T$ on Earth. The entire system (spring and block) is then taken to the Moon, where $g_{Moon} \approx g_{Earth}/6$. A student is asked: *"How does the period of oscillation on the Moon compare to the period on Earth?"*

*For each of the following student responses, answer the two questions below.*

**Student 1 — Anna.** "The period on the Moon is longer. There's less gravity, so the restoring force is weaker, and the block oscillates more slowly." When the teacher asks her to think about it more carefully, she says: "It makes sense — less gravity means everything moves more slowly up there."

a) Why do you think Anna gave this answer?

b) What question would you ask Anna to help her?

**Student 2 — David.** "The period on the Moon is longer. On Earth the equilibrium position is at $mg/k$ below the natural length, but on the Moon it's only $mg/(6k)$. The block is closer to the natural length, so the spring is less stretched at equilibrium and the effective restoring force over a full oscillation is weaker. This means the oscillation is slower and $T_{Moon} > T_{Earth}$."

a) Why do you think David gave this answer?

b) What question would you ask David to help him?

**Student 3 — Emma.** "I'm not sure... maybe the period changes? I know the formula has something to do with mass and the spring constant, but I don't remember whether gravity is in there or not."

a) Why do you think Emma gave this answer?

b) What question would you ask Emma to help her?

Figure 3. The two parallel diagnostic vignette sets, used in a crossover arrangement: (a) set ▲ (coins falling from a table); (b) set ♦ (mass–spring system on the Moon). Each set presents three fictional students whose reasoning instantiates Hazards A+B, C and D. Full texts in the supplemental material.

For each character, participants were asked to diagnose the reasoning and to formulate the question they would ask next. Responses were scored with a rubric on six indicators (diagnosis and proposed question, for each of the three hazards), each on a 0–3 scale, for a total score of 0–18. Diagnosis and question were scored independently: the proposed question was judged against the character's hazard, not against the diagnosis given by the participant.

The simulation logs were coded at the level of the teacher turn along two dimensions: target (which agent, if any, was explicitly named) and strategy (guiding, cognitive conflict, falling-back, or re-explanation). Agent mentions were counted over all 91 teacher turns, while strategy percentages were computed over the 71 substantive turns, excluding greetings, roll calls and comprehension checks. We also tracked a specific pattern in the responses of Davide, the agent embodying Hazards A+B: turns in which he verbally accepted the teacher's point while signalling that his intuition persisted (for example, "ok… but it still feels weird").

This pattern mirrors the dissociation between formal knowledge and intuitive reasoning described by Kryjevskaia et al [11]. We applied a strict criterion: a turn was counted only if it contained both the acceptance and the residual doubt; turns in which Davide expressed doubt while still rejecting the teacher's point were not counted. The occurrence of DPT-specific vocabulary in teacher turns was counted by keyword search, retaining clear cases only.

Two short individual sheets bracketed the simulation, both completed in about five minutes, individually and without discussion. On the *hypotheses sheet*, filled in after reading the problem and the five agents' initial responses but before interacting, participants answered for each simulated student: "Why do you think they gave this response?", "What might be going on in their reasoning?". On the *reflection sheet*, filled in immediately after the session, participants chose one of the five students and answered: "What was wrong with their reasoning?", "What did you try to do to help them? What questions did you ask?", and "Did it work? Why or why not?". Both sheets were coded with the same keyword criterion for DPT vocabulary, and used qualitatively to triangulate episodes in the session logs. The individual sheets and the rubric are available in the supplementary materials.

## 4.4 Analysis

Given the small sample and the ordinal nature of the rubric scores, we used non-parametric tests throughout. Pre–post comparisons used exact two-sided Wilcoxon signed-rank tests, with effect sizes computed as $r = |Z|/\sqrt{N}$ and interpreted following Cohen's conventions; the effect of vignette-set order was checked with exact Mann–Whitney tests. Given the exploratory nature of the pilot, we report p-values without correction for multiple comparisons. Counts from the session logs and from the written sheets are reported descriptively, with no inferential testing. All analyses were run in Python. All scoring and coding were performed by a single rater (the author); we return to this limitation in the Discussion.

# 5. Results

## 5.1 Pre–post gains on the diagnostic vignettes

After the sequence, nearly every participant improved. Nine of the eleven paired participants scored higher at POST, one was stable, and one declined by two points. On the 0–18 scale, the mean total score rose by almost three points, from 13.9 to 16.7 — a gain large enough to be statistically reliable even in this small sample (exact Wilcoxon signed-rank test, $p = 0.014$; effect size $r = 0.79$, large by conventional standards). The change is most tangible at the level of individual responses. Each participant produced six scored responses (a diagnosis and a proposed question for each of the three vignette characters), 66 in all: before the simulation, about half of them earned full marks; afterwards, four in five did — and weak responses (0 or 1 out of 3) all but disappeared, from one in seven to a single one.

The gain was not spread evenly: it concentrated where there was both room and need (table 2). At PRE, participants were already good at recognising the classic wrong intuition — diagnosis scores for Hazards A+B averaged 2.55 out of 3, leaving little space for measurable growth. What grew instead was the ability to formulate the appropriate next question: aggregated across the three characters, proposed-question scores improved significantly (+1.45 points out of 9) while diagnosis scores did not (+1.36, not significant).

In other words, the sequence moved most the component it was designed to move — the questions participants would ask. Looking at single indicators, the largest improvement occurred precisely where participants started lowest: the diagnosis of Hazard C, analytical reasoning recruited to defend an intuitive answer, rose from 1.82 to 2.64 out of 3 (table 2 reports totals per hazard, diagnosis and question combined). Recognising a student who "has a formula" proved harder, at the outset, than recognising a student who "has an intuition" — a point we return to in the Discussion.

**Table 2** Pre–post comparisons (n = 11; exact two-sided Wilcoxon signed-rank tests; effect size r = |Z|/√N).

| Comparison | Scale | Mean gain | p | r |
|---|---|---|---|---|
| **Total score** | 0–18 | +2.82 | 0.014 | 0.79 |
| **Proposed question** (all characters) | 0–9 | +1.45 | 0.031 | 0.91 |
| **Diagnosis** (all characters) | 0–9 | +1.36 | 0.123 | 0.51 |
| **Hazards A+B** | 0–6 | +0.82 | 0.031 | 0.92 |
| **Hazard C** | 0–6 | +1.09 | 0.065 | 0.61 |
| **Hazard D** | 0–6 | +0.91 | 0.027 | 0.78 |

Finally, the crossover behaved as intended: the order in which the two vignette sets were administered had no detectable effect (exact Mann–Whitney tests, all $p > 0.24$), supporting the use of the two sets as equivalent forms. Figure 4 shows each participant's trajectory from PRE to POST: all eleven converge to the upper score band, with no regressions in either panel.

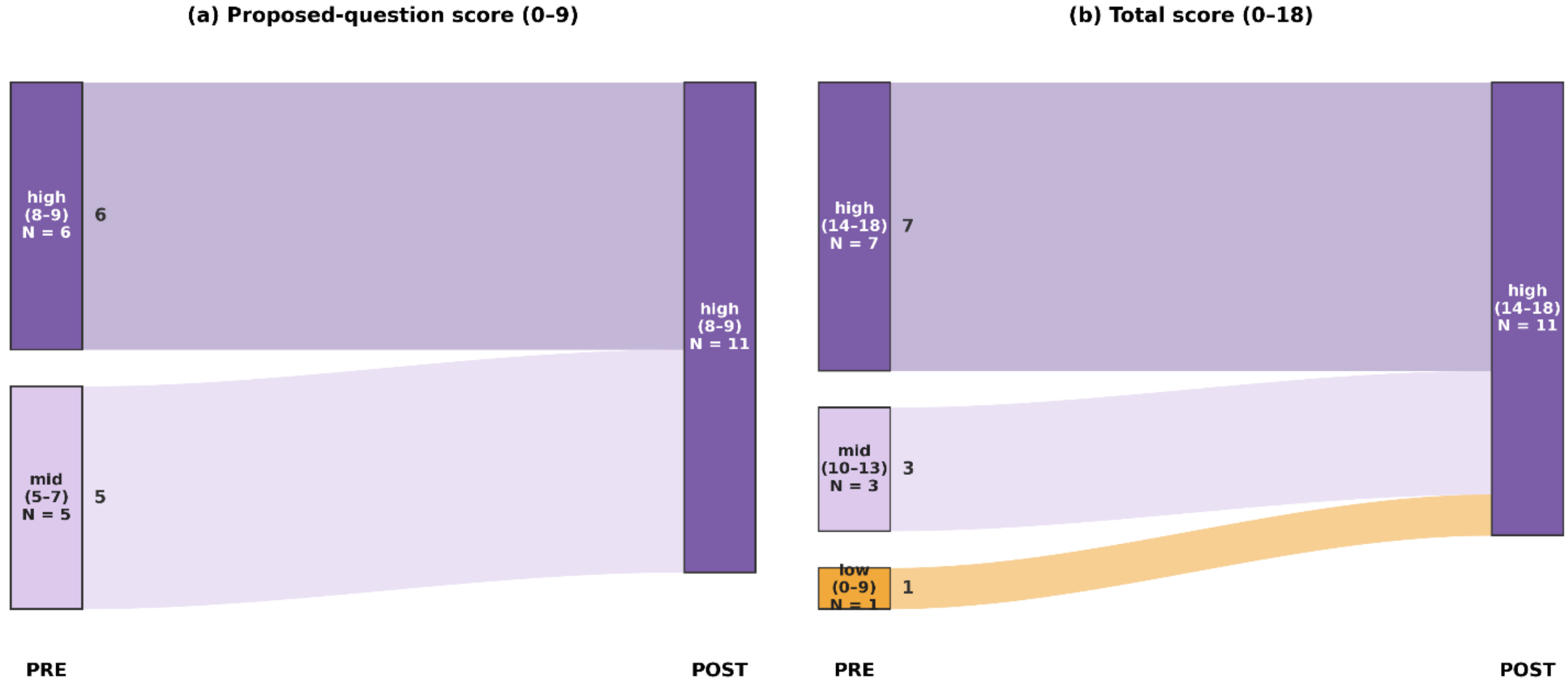

Figure 4. Pre–post trajectories of the n = 11 paired participants, grouped in score bands. (a) Proposed-question score, summed over the three vignette characters (0–9; bands: high 8–9, mid 5–7, low 0–4). (b) Total vignette score (0–18; bands: high 14–18, mid 10–13, low 0–9). Band width is proportional to the number of participants. At POST, all eleven participants reach the upper band on both measures, with no downward transitions.

## 5.2 During the simulation: undifferentiated questioning

The eight simulation sessions lasted between 10 and 42 minutes. Across all 91 teacher turns, the agent embodying Hazards A+B (Davide) was named 24 times — 35% of all named mentions, against a 20% baseline expected from five agents — followed by Elena and Tommaso (14 each), Andrea (9) and Chiara (7). Attention thus gravitated towards the most legible form of student difficulty: a clear wrong claim with intuitive grounding. The individual reflections mirror this: asked to choose one student to write about, 7 of 15 participants chose Davide.

Strategy use was strongly uniform. Of the 71 substantive teacher turns, 49 (69.0%) were guiding questions, 10 (14.1%) cognitive-conflict questions, 10 (14.1%) falling-back questions and 2 re-explanations. The two strategies that the DPT framework prescribes for specific hazards — conflict for entrenched intuition, falling-back for missing knowledge — together accounted for 28.2% of substantive turns, and were rarely matched to the hazard of the agent addressed. Individual sessions, however, spanned a wide range. At one extreme, one participant addressed all five agents in a single coordinated turn with a different question for each; this participant scored 18/18 at POST. At the other, the only session that failed to converge to the correct answer combined exclusive attention to Davide (5 of 9 substantive turns) with a premature endorsement of his error: "Esatto, vedo che hai capito il concetto" ("Exactly, I see you got the concept") immediately followed by Davide restating the incorrect claim.

All other sessions converged. Convergence was accompanied by a recurrent pattern: Davide produced verbal acceptance with a residual-doubt marker ("ok… but it still feels weird") 13 times across seven of the eight sessions. The only session with no such instance was the failed one.

## 5.3 Paper versus practice: the silence of the DPT lexicon

The vocabulary of the framework the participants had been taught — hazard, process 1/2, bias, intuition — appeared in only 2 of the 71 substantive teacher turns. The two occurrences sit at opposite poles of the corpus: in the failed session, as an accusation without a follow-up strategy ("Davide, did you rely purely on intuition?") and in the most differentiated session, as a task ("la tua intuizione fisica" — the participant asks Davide to justify his physical intuition mathematically).

The written sheets show that the vocabulary was available to participants. Already at PRE — completed after the lesson introducing the DPT framework — 5 of the 12 sheets written in class contained DPT terms, and 7 of 13 POST sheets used them, four with explicit hazard labels. The sharpest version of the contrast sits within the simulation day itself: minutes before interacting, 7 of the 14 individual hypotheses sheets described the agents' expected reasoning in DPT terms — in the interaction that followed, the vocabulary appeared in 2 turns of 71. The reflections written minutes after the session point the same way: only 2 of 15 contained DPT terms, and none used hazard labels. The contrast, then, is not between before and after: it tracks distance from the action — the closer participants were to the live exchange, the less they used the framework's language. Nor was it only the vocabulary. On the same hypotheses sheets, most participants had also read the agents correctly: 13 of 14 recognised Davide's intuitive shortcut, 9 Elena's biased use of the formula, 8 Tommaso's knowledge gap — with Davide by far the easiest to read, the same asymmetry seen in the vignettes. In writing, participants knew who was failing and how; in the interaction that followed, they questioned everyone in much the same way.

## 5.4 Participants' perceptions

Responses to the five-item feedback questionnaire (1–5 scale, n = 11) were positive on all items (table 3). The lowest-rated item concerned the realism of the simulated students, consistent with the limitation participants themselves articulated (section 6); one participant also commented that finding productive questions was still difficult, attributing this to being out of practice with the content. The highest-rated items were the recognition that different reasoning errors require different teaching strategies, and the willingness to recommend the simulation for teacher training.

**Table 3.** Feedback questionnaire (1 = strongly disagree, 5 = strongly agree; n = 11). Item wording verbatim from the questionnaire.

| Item | Mean | Range |
|---|---|---|
| 1. The simulation helped me understand how students reason differently | 4.36 | 4–5 |
| 2. The simulation helped me think about what questions to ask students | 4.27 | 2–5 |
| 3. I found the simulated students realistic | 4.09 | 3–5 |
| 4. After this experience, I realize that different reasoning errors require different teaching strategies | 4.55 | 3–5 |
| 5. I would recommend this simulation for teacher training | 4.55 | 3–5 |

# 6. Discussion and limitations

This pilot produced two types of results. The first relates to the design itself: the sequence of experiencing the reasoning hazards first-hand, learning explicit questioning strategies and practising them on a simulated class proved feasible using standard tools. The second type of result is the most instructive for teacher preparation: a dissociation between paper and practice. The central result of this pilot is a dissociation. On paper, participants improved markedly: the total vignette score rose with a large effect ($p = 0.0137$, $r = 0.794$), and the DPT vocabulary came readily to them in writing — already on 5 of the 12 PRE sheets, and on 7 of 13 POST sheets. In real-time interaction, however, the same participants questioned the simulated students in a largely undifferentiated way: guiding questions for everyone, that same vocabulary all but silent (2 turns of 71).

We argue that this gap should not be read as a failure of the instruction or of the participants, but as an accurate snapshot of where they stand on the developmental trajectory of responsiveness to student ideas [13]. Identifying reasoning and responding to it in real time are different competences, and the second develops later — as documented even for written responses [13]. What the simulation adds is the ability to locate each participant on that trajectory with transcript-level granularity, something classroom observation rarely affords.

Three features of the corpus refine this picture. First, the gap is bidirectional: doing sometimes ran ahead of knowing. The clearest case is the fastest-converging session. At PRE, one of the two participants in this session had met Hazard C on the vignette: a character who defends an intuitive answer with seemingly analytical reasoning. The participant found that reasoning convincing — judged the answer correct, and scored zero on that diagnosis. Three days later, in the simulation, the pair met the same kind of reasoning in a formula-based guise — the reading of $\mu_s N$ as the actual friction force — and dismantled it with a single question that stepped back to a simpler case: "If the applied force were zero, how much would the friction be?". In their written reflections immediately after the session, both partners described this move in purely physical terms, with none of the framework's vocabulary: they had found the question, but not yet the words to name what it did. The pattern across all eight sessions points the same way — falling-back questions peak mid-session and cognitive conflict rises towards the end — suggesting strategies discovered in the course of the interaction rather than planned in advance. This is even more remarkable when you consider that each pair had only one session of practice.

Second, possessing a strategy does not guarantee aiming it at the right hazard: in one session a falling-back question was aimed at the rationalising agent rather than the knowledge-gap agent, and the agent answered the prerequisite question correctly — emerging more confident in her wrong answer. The key competence is not just having strategies, but matching strategies to the hazard.

The third feature is that the sequence appears to realign knowing and doing. At PRE, the two halves of a response often came apart: some participants diagnosed a character correctly but then proposed a question that missed the point, while others proposed a good question for a character they had misdiagnosed. At POST, such cases had almost disappeared: diagnosis and question now tended to go together. This movement towards coordination is precisely what Dodlek and colleagues place at the heart of being "attentive and responsive to students' explanations" [13].

The failed session brings these risks together. The participant in this session had recognised the nature of the agent's error — indeed, one of the corpus's two occurrences of DPT vocabulary is this participant naming Davide's intuition explicitly. Yet the diagnosis did not shape the strategy: the questions kept going to Davide alone, and the session closed with the teacher endorsing his wrong answer. In the feedback questionnaire, the same participant nevertheless rated the experience positively on every item.

The corpus also offers a practical cue that future debriefs could make explicit. When a session was going well, Davide's replies changed in a recognisable way: he would accept the teacher's argument while admitting that his intuition still resisted — "ok… but it still feels weird". This happened in seven of the eight sessions, and all seven converged. In the failed session it never happened: Davide kept restating his claim with full confidence. The resulting rule of thumb is simple. If Davide concedes but still hesitates, his analytical thinking is already at work, and guiding questions will move him forward. If he shows no hesitation at all, there is nothing yet for guidance to work on: the teacher first needs to create doubt, for instance with a conflict-type question, and only then guide.

The debrief plausibly played a part in this return. On the day of the simulation itself, the framework's language was almost absent even minutes after the session (2 of 15 reflections); four days later, on the POST vignettes completed right after the class had re-examined excerpts of its own transcripts, it was back on 7 of 13 sheets — and four participants spontaneously classified the new vignette characters by hazard, something no one had done at PRE. In this design the debrief cannot be separated from the four days themselves; but the pattern fits its intended role: the simulation provides the shared experience, the debrief the moment to reconnect it to the framework.

However, the limitations of this pilot must be acknowledged. The paired sample is small (n = 11), powered only for large effects, and the design cannot separate the contributions of simulation, debrief and the surrounding sequence. All scoring and coding were performed by a single rater as disclosed in the Methods. The hazards were reified into stable characters — a pedagogical simplification the framework itself warns against — and the two vignette sets, while parallel, were not formally equated. We note, however, that the POST vignettes involved different characters and different physics content from the simulation, so the measured gains already reflect some transfer beyond the five specific agents.

The lexicon count is keyword-based: it registers the framework's words, not its application. The strategy coding addresses the latter.

Each pair interacted for a single session. Repeated sessions had been planned, but scheduling constraints made them impossible. The log cannot be attributed to individuals: one member of the same pair scored 18/18 at POST while the other showed the only decline. Participants themselves noted the environment's limits: "real students do not ask all those questions".

Finally, the intervention came early in these participants' professional development, before extended practice with responsive teaching; the observed gap is consistent with the developmental literature, and follow-up studies at later stages of the preparation programme would trace how it narrows [13].

# 7. Conclusions

We presented an instructional sequence on dual-process reasoning hazards and questioning strategies. The core of this intervention is a multi-agent AI classroom, that serves as a low-stakes environment, where prospective teachers practise responding with differentiated questioning strategies, in real time, to five simulated students who reason in different ways, from intuitive error to near-expert. In this pilot, participants' diagnostic scores improved with a large effect (n = 11) after the sequence, while their in-simulation questioning remained largely undifferentiated; a knowing-doing gap that the transcript-level record makes visible and locatable for each participant.

Building the simulation required some technical work — a Python application, iterative prompt refinement for the LLM model— but no specialised infrastructure or model training, and once deployed it ran reliably in class. Participants received it positively, while also noting its limits as a proxy for real classrooms.

Future work follows directly. In a next iteration, with a new cohort, the simulation will be placed towards the end of the course, once participants have had extended practice with the questioning repertoire: the expectation, given the developmental framing above, is that later placement should narrow the gap between diagnostic knowledge and enacted questioning. Practice could also extend over multiple sessions: after a first session in class, in pairs, further sessions could be assigned as individual homework. In the spirit of Dodlek, Planinsic and Etkina's comparison of differently prepared teachers, it would also be informative to observe whether teachers experienced in frameworks such as ISLE question the simulated class differently [13].

We hope this architecture, which runs on widely available tools, can activate interest among teacher educators as a low-stakes environment where two entangled skills become visible, and improvable: attending to the reasoning processes behind students' answers, and responding to them with the right question.

# Acknowledgments

The author is grateful to Eugenia Etkina for her advice on physics teacher preparation and for all he learned in her workshops on questioning strategies, which directly shaped the design of this study. The author also thanks Marta Carli for pointing out the core literature on dual-process reasoning.

# Data availability statement

The vignette texts, the individual sheet templates, the scoring rubric and the feedback questionnaire are provided as supplementary material. The simulation, together with the full system prompt defining the five student personas, will be made available on Hugging Face Spaces upon publication. Anonymised session transcripts and further data are available from the author on reasonable request.